\documentclass[11pt,fleqn]{article} 
\usepackage{graphicx}
\usepackage{textcomp} 
\usepackage{epstopdf}
\usepackage{latexsym}
\usepackage{amsmath}
\usepackage{amssymb}
\usepackage{bm}
\usepackage[mathcal]{euscript}
\usepackage{slashed}
\usepackage{xcolor}

\usepackage[top=30mm, bottom=3 0mm, left=25mm, right=25mm]{geometry}

\numberwithin{equation}{section}

\usepackage{marvosym}

\usepackage{indentfirst}

\newcommand\blfootnote[1]{
  \begingroup
  \renewcommand\thefootnote{}\footnote{#1}
  \addtocounter{footnote}{-1}
  \endgroup
}

\usepackage{hyperref}
\hypersetup{
        unicode,	
        colorlinks,
        citecolor=blue,
        linkcolor=blue, 
        urlcolor=red,
        bookmarksopen=true,
        bookmarksopenlevel=\maxdimen,
      }

\def\gl#1#2{\ifmmode \mathrm{GL}(#1; {\bf #2}) \else $\mathrm{GL}(#1; {\bf #2})$\fi}
\def\sl#1#2{\ifmmode \mathrm{SL}(#1; {\bf #2}) \else $\mathrm{SL}(#1; {\bf #2})$\fi}
\def\so#1{\ifmmode \mathrm{SO}({#1}) \else $\mathrm{SO}(#1)$\fi}

\def\sp#1#2{\ifmmode \mathrm{Sp}(#1; {\bf #2}) \else $\mathrm{Sp}(#1; {\bf #2})$\fi}
\def\usp#1{\ifmmode \mathrm{USp}(#1) \else $\mathrm{USp}(#1)$\fi}
\def\spin#1{\ifmmode \mathrm{Spin}(#1) \else $\mathrm{Spin}(#1)$\fi}
\def\su#1{\ifmmode \mathrm{SU}({#1}) \else $\mathrm{SU}(#1)$\fi}

\def\double #1{#1{\hbox{\kern-2pt $#1$}}}

\mathcode`\*="702A                  
\def\half{{\textstyle{1\over{\raise.1ex\hbox{$\scriptstyle{2}$}}}}}

\def \p{\partial}
\def \a{\alpha}
\def \b{\beta}
\def \e{\epsilon}
\def \d{\delta}

\def \g{\gamma}
\def \l{\lambda}

\def \L{\Lambda}

\def \o{\omega}

\def \O{\Omega}

\def\Pib{{\overline\Pi}}
\def\Jb{{\overline J}}
\def\N{{\nabla}}
\def\Nb{{\overline\nabla}}
\def\pb{{\overline\partial}}

\begin{document}

\begin{flushright}
\makebox[0pt][b]{}
\end{flushright}

\vspace{40pt}
\begin{center}
{\LARGE Superspaces for heterotic pure spinor string compactifications
}

\vspace{40pt}
Osvaldo Chandia${}^{\clubsuit}$ and Brenno Carlini Vallilo${}^{\spadesuit}$
\vspace{40pt}

{\em 
${}^{\clubsuit}$ Departamento de Ciencias, Facultad de Artes Liberales \\ Universidad Adolfo Ib\'a\~nez, Chile }\\

\vspace{20pt}

{\em 
${}^{\spadesuit}$ Departamento de Ciencias F\'{\i}sicas, Universidad Andr\'es Bello \\ Sazi\'e 2212, Santiago, Chile}

\vspace{60pt}
{\bf Abstract}
\end{center}
We study supersymmetry conditions for the heterotic pure superstring preserving and $N=1$ supersymmetry in four dimensions directly from the curved superspace defined by the Berkovits-Howe constraints. 

\blfootnote{
${}^{\clubsuit}$ \href{mailto:ochandiaq@gmail.com}{ochandiaq@gmail.com},    
${}^{\spadesuit}$ \href{mailto:vallilo@gmail.com}{vallilo@gmail.com} }

\setcounter{page}0
\thispagestyle{empty}

\newpage

\tableofcontents

\parskip = 0.1in
\section{Introduction}

Understanding string dynamics in curved backgrounds is still one of the most important topics of research in string theory. It is particularly crucial to study dualities and compactifications. Most of the knowledge in string compactifications comes from the Ramond-Neveu-Schwarz (RNS) formalism or the supergravity limit. In the case of heterotic string or Type II strings the RNS formalism allowed significant advances such as non-renormalization theorems \cite{Dixon:1990pc} and exact results from topological strings (see, {\em e.g.}, \cite{Alim:2012gq} for a review).  For compactification backgrounds involving Ramond-Ramond (RR) fields most of the work done was restricted to the supergravity limit. The difficulty comes from the RR vertex operators associated with the linearized RR flux. 

The pure spinor formalism \cite{Berkovits:2000fe} introduced by Berkovits has manifest supersymmetry in flat space-time and made possible to study RR backgrounds without the subtleties of the RR vertex operators of the RNS formalism.\footnote{Another promising  way to study RR backgrounds is to use Closed Superstring Field Theory \cite{deLacroix:2017lif} as in the work \cite{Cho:2018nfn}.} 
In two previous papers \cite{Chandia:2009it,Chandia:2011wd} we started the study of compactifications using the pure spinor formalism\footnote{It was also studied in \cite{Berkovits:2019wgj} using orbifolds}. However, in those works we started with an ansatz for the  covariant super derivative algebra  without relating it the supergravity background defined by the Berkovits-Howe supergravity constraints \cite{Berkovits:2001ue}. The assumption was guided by requiring a nilpotent BRST charge in the supergravity limit. In this work we want to study supergravity backgrounds for the heterotic pure spinor string preserving Poincar\'e symmetry or supersymmetry in four dimensions directly from the curved superspace defined by the Berkovits-Howe constraints. The idea is to construct the Killing supervectors and derive the conditions imposed on the supergeometry such that their lowest components satisfy some general requirements expected for Poincar\'e or supersymmetry parameters. Knowledge of the $\theta$ expansion of Killing supervectors is also useful since they define conserved currents on the worldsheet which can be used to have an explicit form for the supergravity vertex operators.  Killing spinors and their algebra were already studied in, {\em e.g.} \cite{Figueroa-OFarrill:2007omz} but not directly from superspace point of view. 

The superspace approach taken here is considerably more involved that the component approach historically used in the literature on superstring compactifications. However if we want to apply covariant formalisms for the superstring to this problem we must understand all steps in the description of curved superspaces for string compactifications. Although the case studied here is one of the oldest compactification models it will serve the base to apply the same ideas to more general compactifications, {\em e.g.} \cite{Becker:2002sx,Becker:2003yv,Becker:2003sh,Frey:2005zz} and also for Type II strings. Curved superspace methods were also used in \cite{Grassi:2006cd} to describe flux compactifications. 

This paper is organized as follows. In the next section we give a brief review for the heterotic string in the pure spinor formalism in a general curved background and some of the main consequences of the Berkovits-Howe constraints. In the Section \ref{superkilling} we study general Killing supervectors in 
ten dimensional supergravity with the Berkovits-Howe constraints. In Section \ref{solutions} we study the consequences of imposing four dimensional Poincar\'e symmetry and supersymmetry.

\section{Heterotic supergravity and the Berkovits-Howe constraints}
\label{review}
In this section we review the heterotic string in a curved background in the 
 description of the pure spinor formalism.  The world-sheet action is given by
 \begin{align}
S = \int d^2z ~& \left( \frac12 \Pi_{\underline a} \Pib^{\underline a} + \frac12 \Pi^A \Pib^B B_{BA} + d_\a \Pib^\a + \o_\a \Nb \l^\a  + \rho_{\cal A} \N \rho_{\cal A} \right. \cr
 &\left. +d_\a \Jb^I W_I{}^\a + \l^\a \o_\b \Jb^I U_{I\a}{}^\b \right)+ S_{FT} ,
\label{actioncurved}
\end{align}
where the superspace coordinates $Z^M= (X^{\underline m}, \theta^\mu)$ (with ${\underline m}=0, \dots, 9, \mu=1, \dots, 16$) defines $\Pi^A = \p Z^M E_M{}^A(Z) = (\Pi^{\underline a} , \Pi^\a)$, where $E_M{}^A(Z)$ is the vielbein superfield. The world-sheet field $d_\a$ is the generator for superspace translations, $\l^\a$ is the pure spinor variable and $\o_\a$ is its momentum conjugate variable. The world-sheet fields $\rho_{\cal A}$ are the heterotic fermions. They are in the fundamental representation of the gauge group $SO(32)$ or $E_8\times E_8$. These fermions define the current $\Jb^I=\frac12 K^I_{\cal{AB}}\rho_{\cal A}\rho_{\cal B}$, where $K_I$ are the generators of the gauge group Lie algebra in the fundamental representation. The covariant derivatives in (\ref{actioncurved}) are defined by
\begin{align}
\Nb\l^\a=\pb\l^\a+\l^\b \left( \pb Z^M \O_{M\b}{}^\a \right),\quad \N\rho_{\cal A}=\p \rho_{\cal A} + K^I_{\cal{AB}} \left( \p Z^M A_{IM}  \right) \rho_{\cal B} ,
\label{covdev}
\end{align}
where $\O_{M\b}{}^\a$ is the background Lorentz connection and $A_{IM}$ is the gauge group connection. The background fields are $B, W_I, U_I$ and the connections in (\ref{covdev}). Finally, $S_{FT}$ is the Fradkin-Tseytlin term and it is given by
\begin{align}
S_{FT}= \int d^2z ~ \a' r^{(2)} \Phi ,
\label{SFT}
\end{align}
where $ r^{(2)}$ is the world-sheet curvature and $\Phi$ is the dilaton superfield. Note that this term breaks classical conformal invariance but it helps to restore it at the quantum level \cite{Chandia:2003hn}.

The action (\ref{actioncurved}) is the most general expression that is invariant under background Lorentz and gauge transformations such that it is also classically conformal invariant. There is another symmetry that constrains the background fields to satisfy the equations of supergravity in ten dimensions and the equations of super Yang-Mills in a curved background \cite{Berkovits:2001ue}. This symmetry is generated by the pure spinor BRST charge
\begin{align}
Q=\oint dz ~ \l^\a d_\a . 
\label{Qpure}
\end{align}
The nilpotency of $Q$ and the holomorphicity of $ \l^\a d_\a$, imply a set of constraints for the background superfields. These constraints are functions of the  torsion, curvature, field-strength and $H=dB$ components. Let us remind how they are defined. They are given after constructing the super one-forms $E^A=dZ^M E_M{}^A, \O_\a{}^\b=dZ^M \O_{M\a}{}^\b,  A_I=dZ^M A_{IM}$ and the super two-form $B=\frac12 dZ^M dZ^N B_{NM}$. Note that the Lorentz connection has the form 
\begin{align}
\O_\a{}^\b = \d_\a^\b \O^{(s)} + \frac14(\g^{\underline{ab}})_\a{}^\b \O_{\underline{ab}} ,
\label{}
\end{align}
where $\O^{(s)}$ is the connection for scalings and $\O_{\underline{ab}}$ is the usual Lorentz connection in ten-dimensional superspace. In this way, one can define the connection $\O_A{}^B$ with non-zero components $\O_{\underline a}{}^{\underline b}$ and $\O_\a{}^\b$. The matrix $\g^{\underline{ab}}=\frac12(\g^{\underline a}\g^{\underline b}-\g^{\underline b}\g^{\underline a})$, where $(\g^{\underline a})_{\a\b}$ and $(\g^{\underline a})^{\a\b}$ are the symmetric  $16\times16$ $\g$-matrices in ten dimensions that satisfy the Dirac algebra
\begin{align}
(\g^{\underline a})_{\a\g} (\g^{\underline b})^{\g\b} + (\g^{\underline b})_{\a\g} (\g^{\underline a})^{\g\b} = 2 \eta^{\underline{ab}} \d_\a^\b .
\label{dirac}
\end{align} 

The map of a tensor with superspace curved indices to local target space indices is performed with the use of the vielbein and its inverse. For example,
\begin{align}
B_{NM}=(-1)^{N(M+A)} E_M{}^A E_N{}^B B_{BA} ,
\label{btob}
\end{align}
where $(A,M)$ assigns a sign $(+1)$ for bosonic directions and $(-1)$ for fermionic directions (for more details see \cite{Wess:1992cp}). The torsion, curvature and field-strength super two-forms are given by
\begin{align}
&T^A=\N E^A = d E^A + E^B \O_B{}^A,\quad R_B{}^A=d\O_B{}^A + \O_B{}^C \O_C{}^A, \quad F_I= dA_I + f^{JK}{}_I A_J A_K ,\cr
&T^A=\frac12 E^B E^C T_{CB}{}^A,\quad R_B{}^A=\frac12 E^C E^D R_{DCB}{}^A,\quad F_I = \frac12 E^B E^A F_{IAB} ,
\label{TRF}
\end{align}
where the product between forms is the wedge product and $f^{JK}{}_I$ are the structure constants of the gauge group Lie algebra. The covariant derivative $\nabla$ is 
\begin{align}\label{covderivative}
    \nabla = E^A\nabla_A= dZ^M(\partial_M +\frac12 \Omega_M{}^{\underline{ab}} M_{\underline{ab}}+ \Omega^{(s)}_M S),
\end{align}
where $M_{ab}$ are the Lorentz generators and $S$ is the scaling generator. 

Together with
\begin{align}
H=\frac16 E^C E^B E^A H_{ABC} ,
\label{HdB}
\end{align}
the torsion, the curvature and the field-strength satisfy Bianchi identities. They are given by
\begin{align}
&\N_{[A} T_{BC]}{}^D + T_{[AB}{}^E T_{EC]}{}^D - R_{[ABC]}{}^D = 0 ,\cr
&\N_{[A} R_{BC]D}{}^E + T_{[AB}{}^F R_{FC]D}{}^E = 0 ,\cr
&\N_{[A} F_{IBC]}+T_{[AB}{}^D F_{IDC]}= 0 ,\cr
&\N_{[A} H_{BCD]} + \frac32 T_{[AB}{}^E H_{ECD]} = 0 ,
\label{bianchis}
\end{align}
where the (anti)-symmetrization is on $(ABC)$ in the first three equations and it is on $(ABCD)$ in the last equation. 

The nilpotency of the pure spinor BRST charge and the holomorphicity of the BRST current, together with gauge fixing gauge symmetries of the action, imply that some of the torsion components are 
\begin{align}
T_{\a\b}{}^{\underline a}= - (\g^{\underline a})_{\a\b},\quad T_{A\a}{}^\b=0,
\label{tcomp}
\end{align}
some of the $H$ components are
\begin{align}
H_{\a\b\g}=H_{\underline{ab} \a}=0,\quad H_{\a\b\underline a}=-(\g_{\underline a})_{\a\b} ,
\label{hcomp}
\end{align}
and for the field-strength components
\begin{align}
F_{I \a\b} = 0.
\label{fcomp}
\end{align}
Using the Bianchi identities of (\ref{bianchis}) and the torsion components in (\ref{tcomp}), one can prove that the constraints for the curvature are implied. Also, the Bianchi identity involving $\N_{(\a} T_{\b\g)}{}^{\underline a}$ implies that 
\begin{align}
T_{\a \underline{ab}} = 2 (\g_{\underline{ab}}\O^{(s)})_\a . 
\label{t1}
\end{align}
It is important to note that since $\Omega^{(s)}_\alpha\neq0$ one has to be careful when acting with $\nabla_\alpha$ on gamma matrices. For example, 
\begin{align}
    \nabla_\alpha (\gamma^{\underline a})_{\beta\delta}=-2\Omega^{(s)}_\alpha (\gamma^{\underline a})_{\beta\delta}.
\end{align}

 Combining the Bianchi identities involving $\N_{[\a} T_{\underline{ab}]}{}^{\underline c}$ and $\N_{[\a} H_{\b\underline{ab}]}$ one can show that
\begin{align}
T_{\underline{abc}}=-H_{\underline{abc}},\quad \O^{(s)}_{\underline a} = 0. 
\label{t2}
\end{align}
The Bianchi identity involving $\N_{[\a} H_{\underline{abc}]}$ together with the equation $(\g^{\underline b})_{\a\b} T_{\underline{ab}}{}^\b=8\N_{\underline a}\O_\a$ (which is derived from the Bianchi identity involving $\N_{[\underline a} T_{\a\b]}{}^\b$ \cite{Chandia:2003hn}) implies that 
\begin{align}
T_{\underline{ab}}{}^\alpha=-\frac{1}{6}(\gamma^{\underline c}\nabla)^\alpha H_{\underline{abc}} -\frac{4}{3}\nabla_{[\underline a}(\gamma_{\underline b]}\Omega^{(s)})^\alpha .
\label{t3}
\end{align}
Finally, the Bianchi involving $R_{(\a\b\g)}{}^\rho$ implies
\begin{align}
T_{\underline{abc}} = - (\g_{\underline{abc}})^{\a\b} \N_\a \O^{(s)}_\b,
\label{t4}
\end{align}
So we have that 
\begin{align}
        H_{\underline{abc}} = (\g_{\underline{abc}})^{\a\b} \N_\a \O^{(s)}_\b.
\end{align}
Note that $\O^{(s)}_\a=\frac14\N_\a\Phi$ as it is required by ghost number anomaly cancellation \cite{Berkovits:2001ue} (also necessary for vanishing of the beta function at one-loop \cite{Chandia:2003hn}). One can also use the Bianchi identities to find the following expressions for the curvatures
\begin{align}\label{Curvatures}
        &R_{ {\alpha\beta}}{}^{\underline{ab}}=\nabla_{( \alpha} T_{ {\beta})\underline{c}}{}^{\underline b}\eta^{\underline{ca}}+T_{( {\alpha\underline{d}}}{}^{\underline c} T_{ \beta ) \underline{c}}^{  b}\eta^{\underline{da}} +\gamma_{ {\alpha\beta}}^{\underline d} H_{ \underline{d}}{}^{\underline {ab}},\cr
        &R_{\alpha\underline{b}}{}^{\underline{cd}} = -\nabla_{\underline b} T_{ \alpha}{}^{\underline{cd}} - T_{\underline b}{}^{[ {\underline{c}\beta} }\gamma_{ {\beta\alpha}}^{\underline d]},\cr
        &R_{\underline{ab}}{}^{\underline{cd}}=-\frac18 (\gamma^{\underline{cd}})_\alpha{}^\beta \left( \nabla_\beta T_{\underline{ab}}{}^\alpha - T_{\beta[\underline{a}}{}^{\underline e} T_{\underline{b}]\underline{e}}{}^{\alpha} \right),\cr
        &R^{(s)}_{\alpha\beta}=\nabla_{(\alpha}\Omega_{\beta)}=\frac14 \gamma^{\underline a}_{\alpha\beta}\nabla_{\underline a}\Phi,\cr
        &R^{(s)}_{\underline{a}\alpha} = \nabla_{\underline a}\Omega_\alpha=\frac14 \nabla_{\underline a}\nabla_\alpha\Phi,\cr
        &R^{(s)}_{\underline{ab}}=T_{\underline{ab}}{}^\alpha\Omega_\alpha=\frac14 T_{\underline{ab}}{}^\alpha\nabla_\alpha \Phi.
\end{align}

All the gauge covariant background fields depend on the $\Phi$ superfield. This is not something new and it is also true with a different set of constraints \cite{Nilsson:1981bn}. It is interesting to note that for the Type IIB pure spinor string this is not true \cite{Howe:1983sra}. It can be shown that using the Berkovits-Howe constraints the Ramond-Ramond $5$-form, as well as curvature generated from it, does not appear as a higher component of $\Phi$ \cite{ChandiaUnpublished}. An explicit example of this is the $AdS_5\times S^5$ background \cite{Berkovits:2000fe,Chandia:2017afc} where all covariant derivatives of $\Phi$ vanish. 

\section{Killing supervectors}
\label{superkilling}

As discussed in the previous section all Lorentz covariant tensors can be computed from $\Phi$. Under a general super reparametrization $\xi^M(Z)$ it transforms as 
\begin{align}
    \delta \Phi =\xi^M \nabla_M\Phi=\xi^A\nabla_A\Phi=0,
\end{align}
where we are defining $\xi^A=\xi^M E_M{}^A$. However it is not enough to impose that the transformation above vanishes. Generically the covariant tensors are given by 
\begin{align}
    T_{A_1\cdots A_n}\sim \nabla_{A_1}\cdots \nabla_{A_n}\Phi,
\end{align}
therefore in order to have $\delta T_{A_1\cdots A_n}=0$ we must also impose that $\delta\nabla_A=0$.
Another way to see it is that particular components of the vielbein cannot be obtained from $\Phi$ so to impose that the full background is invariant under some particular super diffeomorphism the covariant derivatives (\ref{covderivative}), which depend on $E_A{}^M$, should also be invariant.

The textbook way, see {\em e.g.} \cite{Buchbinder:1998qv}, to study super diffeomorphisms is to introduce a vector superfield containing $\xi^A$, a compensating Lorentz rotation $\Lambda^{\underline{ab}}$ and scale transformation $\sigma$
\begin{align}
        K =\xi^A\nabla_A +\half\Lambda^{\underline{ab}}M_{\underline{ab}}+\sigma S,
\label{Kis}
\end{align}
Although the pure spinor sigma model has two independent local Lorentz symmetries, acting on vectors and spinors separately, a combination of the two is fixed in the process of solving the Berkovits-Howe constraints. The local scale symmetry is also used to fix the dimension zero torsions, however since we are not eliminating the scale connection from $\nabla_\alpha$ we still have to include the scale transformation in $K$. As we will see this will effectively reduce the structure group of symmetry transformations from Lorentz times scale to only Lorentz, as expected. 

A generic tensor superfield $\mathcal{O}^{A_1\cdots A_n}$ transforms under local reparametrizations  and local Lorentz transformations as generated by $K$ as 
\begin{align}
        \delta {\mathcal{O}}^{A_1\cdots A_n} = K \mathcal{O}^{A_1\cdots A_n}.
\end{align}
Since the covariant derivatives map tensors to tensors we have that the covariant derivatives themselves transform as 
\begin{align}
        \delta \nabla_A = [K,\nabla_A].
\end{align}

If we apply this idea to $\nabla_{ \alpha}$ and use the Berkovits-Howe constraints we get 
\begin{align}\label{InvNablaAlpha}
        \delta\nabla_{ \alpha}=&\left(-\nabla_{ \alpha}\xi^{ \beta} {-\frac14}\Lambda^{\underline{ab}}(\gamma_{\underline{ab}})_{ \alpha}{}^{ \beta}-\delta_\alpha^\beta \sigma\right)\nabla_{ \beta} +\left(-\nabla_{ \alpha}\xi^{\underline b} +\xi^{ \beta} \gamma^{\underline b}_{ {\alpha\beta}}-\xi^{  a} T_{ {\underline{a}\alpha}}{}^{\underline b}\right)\nabla_{\underline b} \cr 
        &+\frac{1}{2}\left(-\nabla_{ \alpha} \Lambda^{\underline{ab}}+\xi^C R_{C \alpha}{}^{\underline{ab}}\right) M_{\underline{ab}} +
        \left(\xi^C R^{(s)}_{C\alpha}-\nabla_\alpha \sigma\right) S=0,
\end{align}
where $R^{(s)}_{AB}$ is the scaling curvature and $R_{AB}{}^{ {ab}}$ is the Lorentz curvature. The first term in (\ref{InvNablaAlpha}) defines $\Lambda^{\underline{ab}}$ in terms of $\xi^{ \alpha}$ it also implies that 
\begin{align}
        \nabla_{ \alpha}\xi^{ \alpha}=-16\sigma,\quad (\nabla\gamma^{\underline{abcd}}\xi)=0.
\label{04form}
\end{align}
Before studying the consequences of these conditions let us first analyze 
the second term of (\ref{InvNablaAlpha}), which has the lowest mass dimension. From it we can write an expression for $\xi^{\alpha}$ in terms of $\xi^{\underline a}$ as
\begin{align}        
         \xi^{ \alpha}= \frac1{10}\gamma_{\underline a}^{ {\alpha\beta}}\nabla_{ \beta}\xi^{\underline a}
         +\frac95 \xi^{\underline a} (\gamma_{\underline a} \Omega)^{ \alpha}=\frac1{10}\nabla_{ \beta} \left(\gamma_{\underline a}^{ {\alpha\beta}}\xi^{\underline a}\right) +\frac85 \xi^{\underline a} (\gamma_{\underline a} \Omega)^{ \alpha}.
      \label{xialphais}
       \end{align}

We can also write the fundamental equation for $\xi^{\underline a}$ using the usual ten dimensional gamma matrix identity
\begin{align}
        \gamma_{\underline{b} ( {\beta\delta}}\left(-\nabla_{ \alpha )}\xi^{\underline b}+\xi^{\underline a} T_{ {\alpha}) {\underline a}}{}^{\underline b}\right) =0, 
\end{align} 
it depends only on $\xi^{\underline a}$ and the background. Using the explicit form of $T_{ {\alpha\underline{a}}}{}^{\underline b}$ and the fact that the $\nabla_{ \alpha}$ derivative of 
$\gamma_{\underline{b}{\beta\delta}}$  is not zero, simplifies this equation to just
\begin{align}\label{FundEq}
        \nabla_{( \alpha}\left((\gamma_{\underline b})_{ {\beta\delta})} \xi^{\underline b}\right)=0.
\end{align}
A nice way to summarize this is that the isometries of a general supergravity background with the Berkovits-Howe constraints \cite{Berkovits:2001ue} are generated by a symmetric bi-spinor  satisfying 
\begin{align}
        \lambda^{ \alpha}\lambda^{ \beta}\xi_{ {\alpha\beta}}=0,\quad 
        \nabla_{( \alpha}\xi_{ {\beta\delta})}=0.
\end{align}


Let us see what consequences are obtained by using the conditions (\ref{04form}) and the expression for (\ref{xialphais}) for  $\xi^\a$. Acting with $\N_\a$ on  (\ref{xialphais}) one obtains 
\begin{align}
\N_\a\xi^\a = \frac1{20} \g_{\underline a}^{\a\b} \{ \N_\a , \N_\b \} \xi^{\underline a} + \frac15 \g_{\underline a}^{\a\b} \O_\a \N_\b \xi^{\underline a} + \frac95 (\N_\a\xi^{\underline a}) (\g_{\underline a}\O)^\a + \frac95 \xi^{\underline a} \N_\a ( \g_{\underline a}^{\a\b} \O_\b ) 
\label{nx1}
\end{align}
Using the equation for $\N_\a \xi^{\underline a}$ from (\ref{InvNablaAlpha}), the anticommutator for the for the first term and $\O_\a=\frac14\N_\a\Phi$ one obtains 
\begin{align}
&\N_\a\xi^\a = \frac1{20} \g_{\underline a}^{\a\b} ( \g^{\underline b}_{\a\b} \N_{\underline b} \xi^{\underline a} + \xi^{\underline b} R_{\a\b\underline b}{}^{\underline a} ) + \frac15 \g_{\underline a}^{\a\b} \O_\a ( \xi^\g \g^{\underline a}_{\g\b} + \xi^{\underline b} T_{\b\underline b}{}^{\underline a} ) \cr
&+\frac95 ( \xi^\b \g^{\underline a}_{\b\a} + \xi^{\underline b} T_{\a\underline b}{}^{\underline a} ) (\g_{\underline a}\O)^\a + \frac9{40} \xi^{\underline a} \g_{\underline a}^{\a\b} \{ \N_\a , \N_\b \} \Phi .
\label{nx2}
\end{align}
Now we use the equation for $\N_{\underline a}\xi^{\underline b}$ from (\ref{InvNablaA}) below, the Bianchi identity involving $R_{\a\b\underline b}{}^{\underline a}$ and note that all the terms with two factors of $\O_\a$ vanish (because they form a factor of the form $(\O\g_{\underline a}\O)$), we obtain
\begin{align}
\N_\a\xi^\a = 16 ~ \xi^\alpha \O^{(s)}_\alpha  \Rightarrow \sigma=-\xi^\alpha \O^{(s)}_\alpha.
\label{nx3}
\end{align}
As we mentioned below equation (\ref{Kis})  this condition on $\sigma$ is effectively eliminating the scale connection from $\nabla_\alpha$, reducing the structure group of the symmetry generators. Note that $\Omega_{\underline a}$ already vanishes as a consequence of the Berkovits-Howe constraints. 

The second condition in (\ref{04form})  does not provide information: in fact, using (\ref{xialphais}) one obtains
\begin{align}
&(\g^{\underline{abcd}})_\a{}^\b \N_\b \xi^\a = \frac1{10}(\g^{\underline{abcd}}\g^{\underline e})^{\a\b} \N_\a \N_\b \xi_{\underline e} + \frac15 (\g^{\underline{abcd}}\g^{\underline e})^{\a\b} \O_\a \N_\b \xi_{\underline e} \cr
&+\frac95 (\g^{\underline{abcd}})_\a{}^\b (\N_\b \xi_{\underline e}) (\g^{\underline e}\O)^\a + \frac95 (\g^{\underline{abcd}})_\a{}^\b \xi_{\underline e} \N_\b (\g^{\underline e}\O)^\a .
\label{nx4}
\end{align}
The first term becomes
\begin{align}
-\frac15 (\g^{\underline{abcd}})_\a{}^\b \N_\b \xi^\a - \frac85 (\xi\g^{\underline{abcd}}\O) + \frac95\xi^{[\underline{a}} (\g^{\underline{bcd}]})^{\a\b} \N_\a \O_\b + \frac{54}{5} \xi^{[\underline{a}} (\O\g^{\underline{bcd}]}\O) 
\label{nx5}
\end{align}
after anticommuting the fermionic derivatives, using the equation for $\N_\a\xi_{\underline e}$  derived from (\ref{InvNablaAlpha}) and the Bianchi identity involving $R_{\a\b\underline{e}f}$. For the second term, after using the equation for $\N_\b\xi_{\underline e}$ from  (\ref{InvNablaAlpha}) one obtains that this term is equal to
\begin{align}
-2(\xi\g^{\underline{abcd}}\O) - \frac{18}{5} \xi^{[\underline{a}}(\O\g^{\underline{bcd}]}\O) .
\label{nx6}
\end{align}
For the third term, we use again the equation for $\N_\b\xi_{\underline e}$ to get
\begin{align}
\frac{18}5 (\xi\g^{\underline{abcd}}\O) -\frac{54}5  \xi^{[\underline{a}}(\O\g^{\underline{bcd}]}\O) .
\label{nx7}
\end{align}
For the fourth term we just contract the gamma matrices to obtain
\begin{align}
-\frac95 \xi^{[\underline{a}}(\g^{\underline{bcd}]})^{\a\b} \N_\a \O_\b - \frac{18}5  \xi^{[\underline{a}}(\O\g^{\underline{bcd}]}\O) .
\label{nx8}
\end{align}
Adding (\ref{nx5}), (\ref{nx6}), (\ref{nx7}) and (\ref{nx8}) we obtain that 
\begin{align}
(\g^{\underline{abcd}})_\a{}^\b \N_\b \xi^\a = -\frac15 (\g^{\underline{abcd}})_\a{}^\b \N_\b \xi^\a \Rightarrow (\g^{\underline{abcd}})_\a{}^\b \N_\b \xi^\a = 0.
\label{nx9}
\end{align}
Therefore the second equation in (\ref{04form}) is identically satisfied. 

The vanishing of $\delta\nabla_{\underline a}$ should be implied by $\delta\nabla_{ \alpha}=0$. Nevertheless, it is still useful to have its explicit form 
\begin{align}\label{InvNablaA}
        \delta\nabla_{\underline a}=&\left(-\nabla_{\underline a} \xi^{ \alpha} {-}\xi^{\underline c} T_{\underline{ca}}{}^{ \alpha}\right)\nabla_{ \alpha} +\left(-\nabla_{\underline a}\xi^{\underline b} {-}\xi^{ \alpha}T_{ \alpha\underline a}{}^{\underline b} {-}\xi^{\underline c}T_{\underline{ca}}{}^{\underline b}{-}\Lambda_{\underline a}{}^{\underline b}\right)\nabla_{\underline b}\cr 
        &+
        \frac{1}{2}\left(-\nabla_{\underline a}\Lambda^{\underline{bc}}+\xi^D R_{D\underline a}{}^{\underline{bc}}\right)M_{\underline{bc}} + \left(\xi^C R_{C\underline a}^{(s)}-\nabla_{\underline a}\sigma\right)S=0.
\end{align}

One can use the equations in (\ref{InvNablaAlpha}) to derive each term above. For example, 
consider first $\{ \nabla_\a , \nabla_\b \}\xi^A$, which is equal to 
\begin{align}
\{ \nabla_\a , \nabla_\b \}\xi^A = \g^{\underline a}_{\a\b} \nabla_{\underline a} \xi^A + \xi^B R_{\a\b B}{}^A ,
\label{NNxi}
\end{align}
and use the equations for $\xi$ and $\L$ derived from $\d \nabla_A=0$ in (\ref{InvNablaAlpha}) and (\ref{InvNablaA}). For $\xi^{\underline a}$, (\ref{NNxi}) implies
\begin{align}
&-\frac14\L_{\underline{bc}} [ \g^{\underline{bc}} , \g^{\underline a} ]_{\a\b}  + \xi^{\underline b} \left( \nabla_{(\a} T_{\b)\underline{b}}{}^a + T_{(\a\underline b}{}^{\underline c} T_{\b)\underline{c}}{}^{\underline a} \right)  + \xi^\g \left( 2\g^{\underline a}_{\g(\a}\O_{\b)} + \g^{\underline b}_{\g(\a} T_{\b)\underline{b}}{}^{\underline a} \right) - 2\sigma \g^{\underline a}_{\a\b} \cr
&=-\g^{\underline b}_{\a\b} \L_{\underline b}{}^{\underline a} + \xi^{\underline b} \left( R_{\a\b\underline b}{}^{\underline a} - \g^{\underline c}_{\a\b} T_{\underline{bc}}{}^{\underline a} \right) - \xi^\g \g^{\underline b}_{\a\b} T_{\g\underline b}{}^{\underline a} . 
\label{NNxi0}
\end{align}
The terms with $\L$ cancel after computing the commutator. The term with $\xi^{\underline b}$ is also zero because of the Bianchi identity involving $R_{\a\b\underline b}{}^{\underline a}$. The terms with $\xi^\g$ are equal to
\begin{align}
\xi^\g \left( 2\g^{\underline a}_{\g(\a} \O_{\b)} + \g^{\underline b}_{(\a\b} T_{\g)\underline b}{}^{\underline a}  \right) = - 2 \g^{\underline a}_{\a\b} (\xi^\g \O_\g),
\label{add0}
\end{align}
where we have used the fact that $T_{\b\underline b}{}^{\underline a}=2(\g_{\underline b}{}^{\underline a}\O)_\b$ and the Fierz identity $\g^{\underline b}_{(\g\a} (\g_{\underline b})_{\b)\rho}=0$. The result of (\ref{add0}) has to cancel the term with $\sigma$ in (\ref{NNxi0}). Therefore,
\begin{align}
\sigma = - \xi^\g \O_\g,
\label{add1}
\end{align}
which is a condition already found before (see (\ref{nx3})). 

As we already mentioned before, the scalar superfield $\Phi$ should also be invariant under the transformations generated by $K$. Since $\xi^{  \g} \O_{  \g} = -\sigma$ and $\O_{  \a}=\frac14\N_{  \a}\Phi$, invariance under $K$ implies that 
\begin{align}\label{NablaAPhi}
        \xi^{\underline a} \nabla_{\underline a}\Phi=4\sigma.
\end{align}

Finally, one can calculate the commutator of two different Killing transformations parametrized by 
$(\xi^A_1,\Lambda^{\underline{ab}}_1,\sigma_1)$ and $(\xi^A_2,\Lambda^{\underline{ab}}_2,\sigma_2)$
\begin{align}
K_1=\xi_1^A \nabla_A + \frac12 \L_1^{\underline{ab}} M_{\underline{ab}}+\sigma_1 S,\quad K_2=\xi_2^A \nabla_A + \frac12 \L_2^{\underline{ab}} M_{\underline{ab}}+\sigma_2 S .
\label{}
\end{align}
Both $K$'s leave $\nabla_A$ invariant, so the parameters $\xi$ and $\L$ satisfy the equations the come from $[K , \nabla_A]=0$. The commutator $K_3=[K_1,K_2]$  is then given by
\begin{align}
&K_3=\left( \xi_1^B \nabla_B \xi_2^A - \xi_2^B\nabla_B\xi_1^A -\xi_2^B\xi_1^C T_{CB}{}^A + \xi^{\underline b}_{[1} \L_{2]\underline b}{}^{\underline a} \d_{\underline a}^A + \frac14 (\g_{\underline{ab}})_\a{}^\b \xi^\a_{[1}\L_{2]}^{\underline{ab}} \d_\b^A  \right) \nabla_A   \cr 
&+\frac12 \left( \xi_2^B \xi_1^A R_{AB}{}^{\underline{ab}} + \xi_{[1}^A \nabla_A \L_{2]}^{\underline{ab}} + \L_1^{\underline{[ac}} \L_{2\underline c}{}^{\underline b]} \right)  M_{\underline{ab}}+{ \left( \xi^A_{[1} \N_A\sigma_{2]} \right)S} .
\label{}
\end{align}
Using the equations from (\ref{InvNablaAlpha}) and (\ref{InvNablaA}) for the derivatives of $\xi_1, \xi_2, \L_1, \L_2$ the expression above can be simplified to
\begin{align}
&K_3 = \left( -\xi^{\underline b}_{[1} \xi^\a_{2]} T_{\a\underline b}{}^{\underline a} + \xi^\a_1 \xi^\b_2 \g_{\a\b}^{\underline a} - \xi^{\underline b}_1 \xi^{\underline c}_2 T_{\underline{cb}}{}^{\underline a} \right) \nabla_{\underline a}  - \xi_1^{\underline a}\xi_2^{\underline b} T_{\underline{ab}}{}^\a  \nabla_\a \cr
&+\frac12\left( \xi_1^A \xi_2^B R_{BA}{}^{\underline{ab}} + \L_1^{\underline{ac}} \L_{2\underline c}{}^{\underline b} \right)  M_{\underline{ab}} \cr
&=-\xi_1^B \xi_2^C T_{CB}{}^A \nabla_A +\frac12\left( \xi_1^A \xi_2^B R_{BA}{}^{\underline{ab}} + \L_1^{\underline{ac}} \L_{2\underline c}{}^{\underline b} \right)  M_{\underline{ab}} + \left( \xi^A_{[1} \N_A\sigma_{2]} \right)S
\label{KillingAlgebra}
\end{align}
So we finally obtain that 
\begin{align}
    \xi^A_3= - \xi_1^B \xi_2^C T_{CB}{}^A, \quad \Lambda_3^{\underline{ab}}=\frac12\left( \xi_1^A \xi_2^B R_{BA}{}^{\underline{ab}} - \L_1^{\underline{c}[\underline{a}} \L_{2\underline{c}}{}^{\underline{b}]} \right),\quad\sigma_3= \left( \xi^A_{[1} \N_A\sigma_{2]} \right).
\end{align}

For the heterotic string it is also possible to include a gauge field background. The superfields present in the sigma model are $(\mathbf{A}_B,\mathbf{W}^{ \alpha}, \mathbf{U}, \mathbf{U}^{\underline{ab}})$. They are valued in the Lie algebra of $E_8\times E_8$ or $SO(32)$ and $\mathbf{A}_B$ is the usual gauge superfield. The Berkovits-Howe constraints imply that they can be written in terms of $\mathbf{A}_{ \alpha}$. One could include the gauge superfield in the covariant derivatives together with a compensating gauge transformation in $K$ and include their contributions to (\ref{InvNablaAlpha}) and (\ref{InvNablaA}). However, a gauge background is better described by its dimension $\frac32$ covariant field strength $\mathbf{W}^{ \alpha}$ and its transformation under $K$ is 
\begin{align}
        \delta\mathbf{W}^{ \alpha}= \xi^A\nabla_A \mathbf{W}^{ \alpha} +\frac14\Lambda^{\underline{ab}}\left(\gamma_{ \underline{ab}}\mathbf{W}\right)^{ \alpha},
\end{align}
where the $\nabla_A$ above includes the gauge connection. In what follows we will focus purely on the geometry and will not include a gauge background. 

\subsection{Flat superspace}
In flat space $\nabla_A\Phi=0$, the solution to (\ref{FundEq}) looks like 
\begin{align}\label{10DkillingVec}
       &\xi^{\underline a}=\epsilon^{\underline a} +\Lambda^{\underline{ab}}x_{\underline b} + (\theta\gamma^{\underline a}\eta) - \frac14 (\theta\gamma^{\underline{abc}}\theta) \Lambda_{\underline{bc}} .
\end{align}
where $\epsilon^{\underline a}$ is the translation parameter, $\Lambda^{\underline{ab}}$ is the Lorentz rotation and $\eta^\alpha$ is the supersymmetry parameter. There are no higher order $\theta$ terms in the expansion. One way to see this is to note that the first term in the second line of (\ref{InvNablaAlpha}) in flat space implies that $\Lambda^{\underline{ab}}$ is constant. It also implies that $\nabla_\delta\nabla_\alpha \xi^\beta =0$, which means $\xi^\beta$ is constant in $x$ and that $\nabla_{[\alpha_1}\cdots \nabla_{\alpha_n]}\xi^\beta\big|_{\theta=0}=0$ for $n\geq 2$ so $\xi^\beta$ is at most linear in $\theta$. This is the complete set of isometries of flat ten dimensional superspace. 

We would like to know how the existence of (\ref{10DkillingVec}) implies a flat ten dimensional superspace. First we have to define the basic properties of $\xi^{\underline a}$ in an appropriate way if the superspace is curved. In this case there is no notion of a reparametrization invariant $\theta$ expansion. The higher components of $\xi^{\underline a}$ should be defined as an expansion using the Grassmann odd covariant derivatives. Using the $\Theta$-variable notation of Wess-Bagger \cite{Wess:1992cp} we can represent the $\theta$ expansion of $\xi^{\underline a}$ as
\begin{align}
        \xi^{\underline a}= \xi^{\underline a}\big|_{\theta=0} +\Theta^{ \alpha}(\nabla_{ \alpha}\xi^{\underline a})\big|_{\theta=0} + \frac12 
        \Theta^{ \beta}\Theta^{ \alpha}(\nabla_{[ \alpha}\nabla_{ \beta]}\xi^{\underline a})\big|_{\theta=0}+\cdots
\end{align}
where $\cdots$ are higher order $\Theta$ terms. For flat superspace we could say $\epsilon^{\underline a}$, $\Lambda^{\underline{ab}}$ and $\eta^{ \alpha}$ were constants, but that is a frame dependent notion. For a general curved superspace we will impose that the components of $\xi^{\underline a}$ satisfy 
\begin{align}\label{LambdaDef}
        &(\xi^{\underline b})\big|_{\theta=0}=\epsilon^{\underline b},\quad 
        (\nabla_{\underline a}\epsilon^{\underline b})\big|_{\theta=0}=-\Lambda_{\underline a}{}^{\underline b},\\
        & (\nabla_{ \alpha}\xi^{\underline b})\big|_{\theta=0}= \gamma^{\underline b}_{ {\alpha\beta}}\eta^{ \alpha},\quad (\nabla_{\underline a} \eta^{ \alpha})\big|_{\theta=0}=0,\\
        &(\nabla_{\underline a}(\nabla_{[ \alpha}\nabla_{ \beta]}\xi^{\underline a})\big|_{\theta=0})\big|_{\theta=0}=0,
\end{align}
where $\Lambda_{\underline{ab}}=-\Lambda_{\underline{ba}}$ and also that 
\begin{align}
        (\nabla_{[ \alpha}\nabla_{ \beta]}\xi^{\underline a})\big|_{\theta=0}=-\frac14\gamma_{ {\alpha\beta}}^{\underline{abc}}\Lambda_{\underline{bc}},
\end{align}
is the same $\Lambda$ defined by (\ref{LambdaDef}). Because of the nested $\theta=0$ projections it is difficult to work with the conditions above. However, it is possible to simplify them in the case where the gravitino (and dilatino) vanishes. The bosonic covariant derivative is defined as 
\begin{align}
        \nabla_{\underline a}=E_{\underline a}{}^{\underline m}(x,\theta) \nabla_{\underline m} + E_{ \underline a}{}^\mu(x,\theta) \nabla_\mu=
         (e_{\underline a}{}^{\underline m}(x) +\cdots)\nabla_{\underline m} + (\psi_{ \underline a}{}^\mu(x) +\cdots)\nabla_\mu. 
\end{align} 
where $\cdots$ are higher order $\theta$ terms. Since $\nabla_\mu$ is the only operator that has a $\theta$ derivative, when $\psi_{\underline a}{}^\mu=0$ we can write 
\begin{align}\label{nested}
        (\nabla_{\underline a}({\mathcal O}\big|_{\theta=0}))\big|_{\theta=0}=
        (\nabla_{\underline a}{\mathcal O})\big|_{\theta=0}. 
\end{align}
this allows us to simplify the projections in the conditions for the components of $\xi^{\underline a}$ and use them in (\ref{InvNablaAlpha}) and (\ref{InvNablaA}). In summary, if we want to impose Poincar\'e invariance in the full ten dimensional superspace we require  that the background is invariant under transformations generated by ten linearly independent vectors defined by $\xi^{\underline a}|_{\theta=0}$ satisfying $\nabla_{\underline a}\xi^{\underline b}|_{\theta=0}=-\Lambda_{\underline a}{}^{\underline b}|_{\theta=0}$, with the further conditions that $\Lambda_{(\underline{ab})}=0$ and $\nabla_{\underline c} \Lambda_{\underline{ab}}|_{\theta=0}=0$. It is clear that these conditions in (\ref{InvNablaA}) automatically imply the space is flat and without bosonic torsion. 

\section{Invariance conditions}
\label{solutions}

Before we start let us summarize the results of the previous section. There are only two independent invariance conditions from which all others can be derived, they are 
\begin{align}\label{InvNablaAlphaPhi}
&\xi^\alpha\nabla_\alpha \Phi+\xi^{\underline a}\nabla_{\underline{a}}\Phi=0,\\ \label{DefiningEqXia}
&\nabla_\alpha\xi^{\underline{b}}=\xi^\beta\gamma^{\underline b}_{\alpha\beta}+\frac12\xi_{\underline a}(\gamma^{\underline{ab}})_\alpha{}^\beta\nabla_\beta\Phi.      
\end{align}
We already saw that the second equation can be written purely in terms of $\xi^{\underline a}$ and $\Phi$, the same can be done with the first and it is simply 
\begin{align}
        \gamma_{\underline a}^{\alpha\beta}\nabla_\alpha \xi^{\underline a}\nabla_\beta\Phi +10\xi^{\underline a}\nabla_{\underline a}\Phi=0.
\end{align}
All other equations can be obtained from the ones above and the Bianchi identities. The last equations in both (\ref{InvNablaAlpha}) and (\ref{InvNablaA}) can be obtained from 
\begin{align}
        \nabla_\alpha\left(\xi^{\underline a}\nabla_{\underline a}\Phi-4\sigma\right)=0,\quad \nabla_{\underline a}\left(\xi^\alpha\nabla_\alpha\Phi+4\sigma\right)=0,
\end{align}
using that $\Omega_\alpha=\frac14\nabla_\alpha\Phi$, the scaling curvatures  $R^{(s)}_{AB}$ in (\ref{Curvatures}) and the first lines of (\ref{InvNablaAlpha}) and (\ref{InvNablaA}).

If we can find a set of superfields $(\xi^A,\Lambda^{\underline{ab}},\sigma)$ such that $\delta\nabla_{ \alpha}=0$ the supergravity background will remain invariant. The idea is to find the superspace conditions for these equations to be satisfied for some specific set of parameters $(\xi^A,\Lambda^{\underline{ab}},\sigma)$ corresponding to translations in four dimensions and one global supersymmetry. From now on we will separate vector indices into  four dimensional and six dimensional parts $ \underline{a}=(a,i)$. We will decompose the parameters as 
\begin{align}
    \xi^A=(\xi^\alpha,\xi^a,\xi^i),\quad \Lambda^{\underline{ab}}=(\Lambda^{ab},\Lambda^{ai},\Lambda^{ij}).
\end{align}
Later, when imposing four dimensional supersymmetry we will also use complex indices $(I,\bar I)$ for the internal space. 

\subsection{Four dimensional Poincar\'e symmetry}
Let us first impose only four dimensional Poincar\'e invariance. Of course the conditions on the background are obvious, however it is still instructive to see how they appear from (\ref{InvNablaAlphaPhi}) and (\ref{DefiningEqXia}). It is not possible in this case to fix $\xi^A=(0,\xi^a,0)$ as superfield conditions. This would constrain the full ten-dimensional superspace to be flat. A less restrictive way to do it is to impose 
\begin{align}\label{inicial}
        \xi^A\big|_{\theta=0} = (0,\epsilon^a,0),\quad \Lambda^{ {ab}}|_{\theta=0}=(l^{ab},0,0),
\end{align}
where $\epsilon^a$ is a four dimensional vector and $l^{ab}$ a four dimensional Lorentz rotation. All higher components of the superfields $(\xi^A,\Lambda^{ {ab}})$ should depend only on $(\epsilon^a,l^{ab})$ and the tensors of the geometry. Since we are dealing with a purely bosonic background we will use that 
\begin{align}\label{BosonicBackground}
    \left(\nabla_{\alpha_1}\cdots \nabla_{\alpha_{2n+1}}\Phi\right)\Big|_{\theta=0}=0.
\end{align}

Staring from (\ref{InvNablaAlphaPhi}) we see that the first component of $\Phi$ should independent of the four dimensional coordinates but it is unconstrained on the internal coordinates. 
From using (\ref{InvNablaAlpha}), (\ref{inicial}) and (\ref{BosonicBackground}) we fix the next order in $\theta$ expansion 
\begin{align}\label{PoincareConds}
    &\nabla_\alpha\xi^\beta|_{\theta=0}= -\frac14 l^{ab}(\gamma_{ab})_\alpha{}^\beta,\cr
    &\nabla_\alpha \xi^a|_{\theta=0}=0,\cr
    &\nabla_\alpha\xi^i|_{\theta=0}=0,\cr
    &\nabla_\alpha \Lambda^{ab}|_{\theta=0}=0,\cr
    &\nabla_\alpha \Lambda^{ai}|_{\theta=0}=0,\cr
    &\nabla_\alpha \Lambda^{ij}|_{\theta=0}=0.
\end{align}
To get the second term in the expansion we first use that 
\begin{align}\label{SecondOrdXiA}
    \nabla_\beta\nabla_\alpha \xi^A &= \frac12[\nabla_\beta,\nabla_\alpha]\xi^A +\frac12\{\nabla_\beta,\nabla_\alpha\}\xi^A \cr
    &=\frac12[\nabla_\alpha,\nabla_\beta]\xi^A +\frac12 \gamma^{\underline a}_{\alpha\beta}\nabla_{\underline a} \xi^A +\frac12 \xi^B R_{\alpha\beta B}{}^A
\end{align}
From the second term of (\ref{InvNablaAlpha}) with $A=\underline{a}$ we get 
\begin{align}\label{SecondOrdXia}
    \frac12[\nabla_\beta,\nabla_\alpha]\xi^{\underline a}=&-\frac12 \gamma^{\underline b}_{\alpha\beta}\nabla_{\underline b} \xi^{\underline a} -\frac12 \xi^B R_{\alpha\beta B}{}^{\underline a}+\nabla_\beta \xi^\delta\gamma_{\alpha\delta}^{\underline a} \cr &+2\xi^\delta\Omega_\beta\gamma_{\alpha\delta}^{\underline a}-\nabla_\beta\xi^{\underline b}T_{\underline{b}\alpha}{}^{\underline a} -\xi^{\underline b}\nabla_\beta T_{\underline{b}\alpha}{}^{\underline a}.
\end{align}

Before projecting to $\theta=0$ we must find the consequences of Poincar\'e symmetry conditions $\nabla_a\xi^b|_{\theta=0}=-l_a{}^b$, $\nabla_a\Lambda^{bc}|_{\theta=0}=0$ from (\ref{InvNablaA}). The first term in (\ref{InvNablaA}) gives no information at leading order in $\theta$. Choosing the indices in the second term to be $(a,b)$ we find that $T_{abc}|_{\theta=0}=0$. From the choice $(a,i)$ and using that $(\nabla_a\xi^i)|_{\theta=0}=0$ we have that $T_{cai}|_{\theta=0}=0$. On the other hand, this will imply that $(\nabla_i\xi^b)|_{\theta=0}=0$, as expected. Finally, from the choice $(i,j)$ we have that $T_{cij}|_{\theta=0}=0$. From the third term in (\ref{InvNablaA}) we get a vanishing four dimensional curvature $R_{ab}{}^{cd}|_{\theta=0}=0$ and also that $R_{ab}{}^{cj}|_{\theta=0}=R_{ab}{}^{ij}|_{\theta=0}=0$. Using Bianchi identities we have that $\nabla_i\Lambda^{cd}|_{\theta=0}$ and $R_{di}{}^{bc}|_{\theta=0}=0$. Using all this information we get the third term in the covariant theta expansion of $\xi^a$ and $\xi^i$
\begin{align}
        \frac12[\nabla_\beta,\nabla_\alpha]\xi^{a}|_{\theta=0}&=\left(\frac12 \gamma^{ b}_{\alpha\beta}l_b{}^a-\frac12 \epsilon^b R_{\alpha\beta b}{}^{ a}-\frac14(\gamma^{cd})_\beta{}^\delta\gamma_{\alpha\delta}^{ a}l_{cd} -\epsilon^{ b}\nabla_\beta T_{{b}\alpha}{}^{ a}\right)\Big|_{\theta=0}\cr
        &= \frac14\gamma^{acd}_{\alpha\beta}l_{cd} -\frac12\epsilon^c \left(\nabla_{[\alpha} T_{\beta]c}{}^a\right)|_{\theta=0}  ,\\
        \frac12[\nabla_\beta,\nabla_\alpha]\xi^{i}|_{\theta=0}&=\frac14\gamma^{icd}_{\alpha\beta} l_{cd}-\frac12\epsilon^c \left(\nabla_{[\alpha} T_{\beta]c}{}^i\right)|_{\theta=0}.
\end{align} 

The next steps would be to calculate $[\nabla_\alpha,\nabla_\beta]\Lambda^{ab}$ and $\nabla_{[\alpha}\nabla_\beta\nabla_{\delta]}\xi^\sigma$.  For the latter we first calculate $[\nabla_\beta,\nabla_\alpha]\xi^\sigma$ using the same method as above and then applying another covariant derivative and anti symmetrizing in all indices. This will give terms depending on the curvature. 
Additional constraints on the background from imposing Poincar\'e symmetry in four dimensions will come from (\ref{InvNablaAlphaPhi}). At lowest order in $\theta$ we only have that 
\begin{align}
        (\xi^\alpha\nabla_\alpha \Phi+\xi^{\underline a}\nabla_{\underline{a}}\Phi)|_{\theta=0} =\epsilon^a(\nabla_a\Phi)|_{\theta=0}=0, 
\end{align} 
so the first component of $\Phi$, the dilaton, is constant in the four dimensional variables. For the next order we have to compute 
\begin{align}
        \nabla_{[\alpha}\nabla_{\beta]} \left(  \xi^\gamma\nabla_\gamma\Phi +\xi^{\underline a}\nabla_{\underline a}\Phi \right)\Big|_{\theta=0}=0.
\end{align}
Using the conditions on the symmetry parameters and evaluating at $\theta=0$ this expression becomes 
\begin{align}
\left( 2(\N_\b\xi^\g) \N_\a\N_\g\Phi-2(\N_\a\xi^\g)\N_\b\N_\g\Phi+([\N_\a,\N_\b]\xi^{\underline a}) \N_{\underline a}\Phi+\xi^{\underline a}[\N_\a,\N_\b]\N_{\underline a}\Phi\right)\Big|_{\theta=0} =0 .
\label{}
\end{align}
Using (\ref{PoincareConds}), (\ref{SecondOrdXiA}) and (\ref{SecondOrdXia}) we simplify to
\begin{align}
\left(\frac12 l_{ab}(\g^{ab})_{[\a}{}^\g (\N_{\b]}\N_{\g}\Phi)-\frac12 l_{ab}\g^{ab\underline c}_{\a\b} \N_{\underline c}\Phi + \e^b \left( \N_{[\a} T_{\b]b}{}^{\underline a}\right)\N_{\underline a}\Phi + \e^a [\N_\a,\N_\b]\N_a\Phi\right)\Big|_{\theta=0}=0.
\label{4.16}
\end{align}
Commuting the derivatives in the last term we obtain
\begin{align}
[\N_\a,\N_\b]\N_a \Phi|_{\theta=0}=-\left(\N_{[\a} T_{\b]a}{}^{\underline b} \right)\N_{\underline b}\Phi|_{\theta=0} + \N_a[\N_\a,\N_\b]\Phi|_{\theta=0},
\label{}
\end{align}
plugging this into (\ref{4.16}), it becomes
\begin{align}
\frac12 l_{ab}(\g^{ab})_{[\a}{}^\g (\N_{\b]}\N_{\g}\Phi)|_{\theta=0}-\frac12 l_{ab}\g^{ab\underline c}_{\a\b} \N_{\underline c}\Phi|_{\theta=0} + \N_a[\N_\a,\N_\b]\Phi|_{\theta=0} =0.
\label{}
\end{align}
Now we use the relation 
\begin{align}\label{NablaNablaPhi}
\N_\a\N_\b\Phi=\frac12\g^{\underline a}_{\a\b}\N_{\underline a} \Phi-\frac1{24}\g^{\underline{abc}}_{\a\b} H_{\underline{abc}}=\frac12\g^{\underline a}_{\a\b}\N_{\underline a} \Phi-\frac1{24}\g^{ijk}_{\a\b} H_{ijk},
\end{align}
to finally obtain
\begin{align}
\left(\frac14 l_{ab}\{\g^{ab},\g^{\underline c}\}_{\a\b}\N_{\underline c}\Phi  + \frac1{48} l_{ab} H_{ijk}[\g^{ab},\g^{ijk}]_{\a\b}-\frac12(\g^{ab\underline c})_{\a\b}\N_{\underline c}\Phi -\frac1{12}\g^{ijk}\N_a H_{ijk} \right)\Big|_{\theta=0} = 0 . 
\label{}
\end{align}
The first term cancels the third term. The second term is vanishes because $\g^{ab}$ commutes with $\g^{ijk}$. Finally we obtain the equation
\begin{align}
(\N_a H_{ijk})|_{\theta=0} = 0,
\label{}
\end{align}
which is the expected condition at mass dimension one from (\ref{InvNablaAlphaPhi}). The calculation at fourth order in $\theta$'s is significantly more involved and will not be presented here

\subsection{Global four dimensional $N=1$ supersymmetry}

We now turn to the conditions imposed by four dimensional global supersymmetry and the calculation of the corresponding Killing supervector. First we want to explain the notation we will use. In the breaking of $SO(1,9)$ to $SO(1,3)\times SO(6)$ the sixteen component spinors will factorize into $SL(2,{\mathbb C})\times SU(4)$ spinors  
\begin{align}
        \mathbf{16}\to ({\mathbf 2},{\mathbf 4})+ (\bar{\mathbf 2},\bar{\mathbf 4}).
\end{align}
The global supersymmetry parameters are the spinors of $SL(2,{\mathbb C})$. All the spinors considered in this section will be assumed to be a Grassman  odd $SL(2,{\mathbb C})$ spinor times a Grassmann even $SU(4)$ spinor
\begin{align}\label{etachi}
        \tilde\eta = \varepsilon\otimes \chi + \bar\varepsilon\otimes\bar\chi.
\end{align}
Will also assume that $\chi$ is normalizable such that $\chi\bar\chi=1$.
Furthermore, when considering a different spinor $\eta'$ we will assume that only the $SL(2,{\mathbb C})$ to be different
\begin{align}
        \tilde\eta' = \varepsilon'\otimes \chi + \bar\varepsilon'\otimes\bar\chi.
\end{align}
This will, for example, imply that 
\begin{align}
        \eta^\alpha\gamma^a_{\alpha\beta}\eta'^\beta= (\varepsilon\gamma^a\varepsilon'),\quad 
        \eta^\alpha\gamma^i_{\alpha\beta}\eta'^\beta=0.
\end{align}

When we assumed that $\chi$ is normalizable it was already implicit that $\chi$ is a nowhere vanishing spinor of the internal manifold which means the internal manifold has $SU(3)$ structure. 
Locally we can choose a tangent space basis such that an $\mathfrak{su}(3)$ subalgebra of $so(1,9)$ annihilates $\eta$. To make that explicitly we will choose complex tangent space indices for the internal manifold $(I,\bar I)$. The antisymmetric product of two gamma matrices can be decomposed as 
\begin{align}
        \gamma^{\underline{ab}}=(\gamma^{ab},\gamma^{aI},\gamma^{a\bar I},\gamma^{IJ},\gamma^{\bar I\bar J},\gamma^{I\bar J},\gamma_{\mathbf 1}).
\end{align}
The spinor $\mathfrak{su}(3)$ generators are $\gamma^{I\bar J}$ and any spinor of the type (\ref{etachi}) will satisfy
\begin{align}
        (\gamma^{I\bar J})_\alpha{}^\beta\tilde\eta^\alpha=0.
\end{align}
It is clear that we can decompose the real spinor $\tilde\eta$ as 
\begin{align}
        \tilde\eta^\alpha=  \eta^\alpha+\bar\eta^{\alpha}.
\end{align}
We will choose the normalization of $\gamma_{\mathbf 1}$ such that 
\begin{align}
        \gamma_{\mathbf 1}\eta =\eta,\quad \gamma_{\mathbf 1}\bar\eta=-\bar\eta.  
\end{align}
Finally, the complex pair of spinors $(\eta,\bar\eta)$ also satisfy 
\begin{align}\label{singlets}
        \gamma^{\bar I}\eta=0,\quad \gamma^I\bar\eta=0. 
\end{align}

With all these set up the Killing supervector and local Lorentz transformation corresponding to $N=1$ supersymmetry transformations 
will satisfy
\begin{align}\label{InitialCondSusy}
        \xi^A|_{\theta=0}=(\tilde\eta^\alpha,0,0),\quad \Lambda^{ {ab}}|_{\theta=0}=(0,0,0),\quad 
        \sigma|_{\theta=0}=0.
\end{align}
Again we can go through the equations imposed by (\ref{InvNablaAlpha}) and (\ref{InvNablaA}) finding the higher components of $(\xi^A,\Lambda^{ {ab}})$ and the conditions imposed in the geometry. In this case the only consequences from the $\theta=0$ projection coming from (\ref{InvNablaA}) are 
\begin{align}\label{covconst}
        (\nabla_{\underline a} \tilde\eta^\alpha)|_{\theta=0}=0
\end{align} 
It is well known (see, {\em e.g.}, \cite{Grana:2005jc,Koerber:2010bx}) that this means it is always possible to choose a connection with $SU(3)$ holonomy. However it does not yet imply Ricci flatness since the torsion is not constrained. Furthermore, it is possible to have curvature terms like $R_{ab}{}^{I\bar J}|_{\theta=0}$ with $\delta_{I\bar J}R_{ab}{}^{I\bar J}|_{\theta=0}=0$. The next order in the $\theta$ expansion is given by (\ref{InvNablaAlpha})
\begin{align}
        &\nabla_\alpha\xi^\beta|_{\theta=0}=0,\\
        &\nabla_\alpha\xi^{\underline a}|_{\theta=0}=\gamma^{\underline a}_{\alpha\beta}\tilde\eta^\beta,\\
        &\nabla_\alpha\Lambda^{\underline{ab}}|_{\theta=0}=\tilde\eta^{\beta}\left(\nabla_{(\alpha}T_{\beta)\underline c}{}^{\underline b}\Big|_{\theta=0}\eta^{\underline{ca}}+\gamma^{\underline c}_{\alpha\beta}H_{\underline c}{}^{\underline{ab}}\Big|_{\theta=0}\right),\\
        &\nabla_\alpha\sigma|_{\theta=0}=\frac14 \tilde\eta^\beta\gamma^{\underline a}_{\alpha\beta}\nabla_{\underline a}\Phi|_{\theta=0}. 
\end{align}

With this information we can compute the first non-trivial supersymmetry condition from (\ref{InvNablaAlphaPhi}), which is the supersymmetry transformation for the dilatino
\begin{align}
        \left(-\xi^\alpha\nabla_\beta\nabla_\alpha\Phi+(\nabla_\beta\xi^{\underline a})\nabla_{\underline a}\Phi\right)\Big|_{\theta=0}=0.
\end{align}
Using (\ref{NablaNablaPhi}) we get 
\begin{align}\label{dilatinoSusy}
        \tilde\eta^\alpha\left(\gamma_{\beta\alpha}^{\underline a}\nabla_{\underline a}\Phi+\frac{1}{12}\gamma_{\beta\alpha}^{\underline{abc}}H_{\underline{abc}}\right)\Big|_{\theta=0}=0.
\end{align}
The different sign from what is usually obtained from the supersymmetry transformation for the dilatino comes from (\ref{t2}). 
At the next order we have to compute $\nabla_{[\beta}\nabla_{\gamma}\nabla_{\delta]}\left( \xi^\a \N_\a \Phi + \xi^{\underline a} \N_{\underline a} \Phi \right)|_{\theta=0}
$. This a long and tedious computation. To simplify it we will restrict to the cases where $\nabla_{\underline{a}}\Phi|_{\theta=0}$ vanishes. First, acting with $\N$'s we get
\begin{align}
&\N_\b \N_\g \N_\d \left( \xi^\a \N_\a \Phi + \xi^{\underline a} \N_{\underline a} \Phi \right)|_{\theta=0} = \left[ -\xi^\a \N_\b\N_\g\N_\d\N_\a\Phi \right. \cr
&-\left(\N_\g\N_\d\xi^\a\right) \N_\b \N_\a \Phi  + \left(\N_\b\N_\d\xi^\a\right) \N_\g \N_\a \Phi - \left(\N_\b\N_\g\xi^\a\right) \N_\d \N_\a \Phi  \cr
&\left. +\left(\N_\d\xi^{\underline a}\right)\N_\b\N_\g\N_{\underline a}\Phi-\left(\N_\g\xi^{\underline a}\right)\N_\b\N_\d\N_{\underline a}\Phi+\left(\N_\b\xi^{\underline a}\right)\N_\g\N_\d\N_{\underline a}\Phi \right]_{\theta=0} ,
\label{3rdOrd}
\end{align}
here we used (\ref{BosonicBackground}). Note that the terms with an expression like $\N_\b\N_{\underline a}\Phi$ go away because $\N_\b\N_{\underline a}\Phi=[\N_\b,\N_{\underline a}]\Phi+\N_{\underline a}\N_\b\Phi$ vanishes at $\theta=0$. Use something similar with the terms with $\N_\b\N_\g \N_{\underline a} \Phi$,
\begin{align}
&\N_\b\N_\g \N_{\underline a} \Phi |_{\theta=0} = \N_\b \left( [\N_\g,\N_{\underline a}]\Phi + \N_{\underline a} \N_\g \Phi \right) |_{\theta=0} =  \N_\b \left( -T_{\g \underline a}{}^{\underline b} \N_{\underline b} \Phi + \N_{\underline a} \N_\g \Phi \right) |_{\theta=0} \cr
&=\left( [\N_\b , \N_{\underline a} ]\N_\g\Phi + \N_{\underline a}\N_\b\N_\g\Phi  \right) |_{\theta=0} = \N_{\underline a}\N_\b\N_\g\Phi |_{\theta=0} ,
\label{}
\end{align}
and (\ref{3rdOrd}) becomes,
\begin{align}
&\N_\b \N_\g \N_\d \left( \xi^\a \N_\a \Phi + \xi^{\underline a} \N_{\underline a} \Phi \right)|_{\theta=0} = \left[ -\xi^\a \N_\b\N_\g\N_\d\N_\a\Phi \right. \cr
&-\left(\N_\g\N_\d\xi^\a\right) \N_\b \N_\a \Phi  + \left(\N_\b\N_\d\xi^\a\right) \N_\g \N_\a \Phi - \left(\N_\b\N_\g\xi^\a\right) \N_\d \N_\a \Phi  \cr
&\left. +\left(\N_\d\xi^{\underline a}\right)\N_{\underline a}\N_\b\N_\g\Phi-\left(\N_\g\xi^{\underline a}\right)\N_{\underline a}\N_\b\N_\d\Phi+\left(\N_\b\xi^{\underline a}\right)\N_{\underline a}\N_\g\N_\d\Phi \right]_{\theta=0} ,
\label{3rdOrdDistribute}
\end{align}
The first term in the rhs of (\ref{3rdOrdDistribute}) is
\begin{align}
&-\frac1{32}(\tilde\eta\g^{\underline{abc}})_\d (\g^{\underline{de}}\g_{\underline a})_{\b\g} R_{\underline{bcde}} - \frac1{24\times 48} (\tilde\eta\g^{\underline{abc}})_\d \g^{\underline{def}}_{\b\g} H_{\underline{abc}} H_{\underline{def}} \cr
&+\frac1{24\times16} (\tilde\eta\g^{\underline{abc}})_\d (\g^{\underline{efg}}\g_{\underline{ad}})_{\b\g} H_{\underline{bc}}{}^{\underline d} H_{\underline{efg}} .
\label{}
\end{align}
The last line in (\ref{3rdOrdDistribute}) is 
\begin{align}
-\frac1{24} (\tilde\eta\g^{\underline a})_\d \g^{\underline{bcd}}_{\b\g} \N_{\underline a} H_{\underline{bcd}} + \frac1{24} (\tilde\eta\g^{\underline a})_\g \g^{\underline{bcd}}_{\b\d} \N_{\underline a} H_{\underline{bcd}} - \frac1{24} (\tilde\eta\g^{\underline a})_\b \g^{\underline{bcd}}_{\g\d} \N_{\underline a} H_{\underline{bcd}} .
\label{}
\end{align}
Using (\ref{NablaNablaPhi}) we obtain
\begin{align}
&\N_\a \N_\b \xi^\g|_{\theta=0} =  - \frac3{16} (\tilde\eta\g^c)_\a (\g^{\underline{ab}})_\b{}^\g H_{\underline{abc}} + \frac1{96} (\tilde\eta\g^{\underline{abc}})_\a \d_\b^\g H_{\underline{abc}}  - \frac1{96}(\tilde\eta\g_{\underline{abcde}})_\a (\g^{\underline{ab}})_\b{}^\g H^{\underline{cde}} .
\label{}
\end{align}
Using this, the second line in (\ref{3rdOrdDistribute} )is equal to
\begin{align}
&\frac1{8\times16} (\tilde\eta\g^{\underline a})_\g (\g^{\underline{bcd}}\g^{\underline{ef}})_{\b\d} H_{\underline{aef}} H_{\underline{bcd}} + \frac3{32} (\tilde\eta\g^{\underline a})_\b (\g^{\underline{bcd}})_{\g\d} H_{\underline{bc}}{}^{\underline e} H_{\underline{eda}} \cr
&+\frac1{96\times24}(\tilde\eta\g^{\underline{abc}})_\g (\g^{\underline{def}})_{\b\d} H_{\underline{abc}} H_{\underline{def}} - \frac1{96\times12}(\tilde\eta\g^{\underline{abc}})_\b (\g^{\underline{def}})_{\g\d} H_{\underline{abc}} H_{\underline{def}} \cr
&+\frac1{96\times24}(\tilde\eta\g_{\underline{abcde}})_\g (\g^{\underline{fgh}}\g^{\underline{ab}})_{\b\d} H^{\underline{cde}} H_{\underline{fgh}} - \frac1{24\times8} (\tilde\eta\g_{\underline{abcde}})_\b (\g^{\underline{agh}})_{\g\d} H^{\underline{cde}} H_{\underline{gh}}{}^{\underline b} .
\label{}
\end{align}
Here we have used the identity $[\g_{\underline{ab}},\g_{\underline{cde}}]=2 \left( \eta_{\underline{a}[\underline{c}} \g_{\underline{de}]\underline{b}} - \eta_{\underline{b}[\underline{c}} \g_{\underline{de}]\underline{a}} \right)$. Using all this we obtain
\begin{align}
&6\N_{[\b}\N_\g\N_{\d]} \left( \xi^\a \N_\a \Phi + \xi^a \N_{\underline a} \Phi \right)|_{\theta=0} = -\frac1{16}(\tilde\eta\g^{\underline{abc}})_{\{\b} (\g_{\underline{ade}})_{\g\d\}} R_{\underline{bc}}{}^{\underline{de}} - \frac1{24\times24}(\tilde\eta\g^{\underline{abc}})_{\{\b} (\g^{\underline{def}})_{\g\d\}} H_{\underline{abc}} H_{\underline{def}} \cr
&-\frac1{64}(\tilde\eta\g^{\underline{abc}})_{\{\b} (\g_{\underline{fgd}})_{\g\d\}} H_{\underline{bc}}{}^{\underline d} H_{\underline a}{}^{\underline{fg}} + \frac1{64}(\tilde\eta\g^{\underline{abc}})_{\{\b} (\g_{\underline{fga}})_{\g\d\}} H_{\underline{bce}} H^{\underline{efg}} \cr
&-\frac14 (\tilde\eta\g^{\underline a})_{\{\b}(\g^{\underline{bcd}})_{\g\d\}} \N_{\underline a} H_{\underline{bcd}} - \frac3{32} (\tilde\eta\g^{\underline a})_{\{\b}(\g^{\underline{bcd}})_{\g\d\}} H_{\underline{ad}}{}^{\underline e} H_{\underline{ebc}} +\frac3{16} (\tilde\eta\g^{\underline a})_{\{\b} (\g^{\underline{bcd}})_{\g\d\}} H_{\underline{bc}}{}^{\underline e} H_{\underline{eda}} \cr
&-\frac1{96\times4} (\tilde\eta\g^{\underline{abc}})_{\{\b}(\g^{\underline{def}})_{\g\d\}} H_{\underline{aef}} H_{\underline{bcd}} - \frac1{96\times2} (\tilde\eta\g_{\underline{abcde}})_{\{\b}(\g^{\underline{agh}})_{\g\d\}} H^{\underline{cde}} H_{\underline{gh}}{}^{\underline b} \cr
&-\frac1{24\times4}(\tilde\eta\g_{\underline{abcde}})_{\{\b} (\g^{\underline{agh}})_{\g\d\}} H^{\underline{cde}} H_{\underline{gh}}{}^{\underline b} = 0.
\label{eqp}
\end{align}
where $\{\a\b\g\}$ means cyclic permutation. The equation (\ref{eqp}) can be simplified to
\begin{align}
&
(\tilde\eta\g^{\underline a})_{\{\b}(\g^{\underline{bcd}})_{\g\d\}} \left( 4 \N_{\underline a} H_{\underline{bcd}} + \frac92 H_{\underline{ad}}{}^{\underline e} H_{\underline{ebc}} \right) \cr
&+(\tilde\eta\g^{\underline{abc}})_{\{\b} (\g^{\underline{def}})_{\g\d\}}  \left(\eta_{\underline{ad}} R_{\underline{bc}}{}_{\underline{ef}} +\frac1{36} H_{\underline{abc}} H_{\underline{def}} + \frac7{24}  H_{\underline{ade}} H_{\underline{bcf}} - \frac14 \eta_{\underline{ad}} H_{\underline{bc}}{}^{\underline g} H_{\underline{gef}} \right) \cr
&+\frac14 (\tilde\eta\g_{\underline{abcde}})_{\{\b} (\g^{\underline{agh}})_{\g\d\}} H^{\underline{cde}} H_{\underline{gh}}{}^{\underline b} = 0. 
\label{lasteq}
\end{align}

So far the conditions found only imply there exists a global Grassmann odd symmetry generated by (\ref{InitialCondSusy}). In order to have $N=1$ supersymmetry we must impose that at zero order in $\theta$ the commutator of two Killing vectors with different parameters in  (\ref{InitialCondSusy}) generate only a four dimensional translation. From (\ref{KillingAlgebra}) we find 
\begin{align}
        &\tilde\eta_1^\alpha\tilde\eta_2^\beta\left(\nabla_{(\alpha}T_{\beta)\underline a}{}^{\underline b} +\gamma_{\alpha\beta}^{\underline c}H_{\underline{ca}}{}^{\underline b}\right)|_{\theta=0}=0,\\
        &\tilde\eta_1^\alpha\tilde\eta_2^\beta\gamma_{\alpha\beta}^{\underline a}\nabla_{\underline a}\Phi|_{\theta=0}=0.
\end{align}
The second equation implies that the dilaton is constant in the four dimensional variables. The first can be written as 
\begin{align}\label{closedSusy}
        \tilde\eta_1^\alpha\tilde\eta_2^\beta\left( 2 \gamma_{\alpha\beta}^{[\underline a}\eta^{\underline{b}]i}\nabla_{i}\Phi + \frac16 \gamma_{\alpha\beta}^{\underline{abcde}}H_{\underline{cde}} \right)\Big|_{\theta=0}=0.
\end{align}
This equation can be combined with one obtained from (\ref{dilatinoSusy}) with a spinor $\tilde\eta_1$ multiplying it by $\tilde\eta_2^\delta (\gamma^{\underline{de}})_\delta{}^\beta$ and then subtracting it from the same equation but with the order of spinors reversed we get 
\begin{align}
       \tilde\eta_1^\gamma\tilde\eta_2^\alpha\left( 2[\gamma^{\underline{de}},\gamma^{\underline a}]_{\alpha\gamma}\nabla_{\underline{a}}\Phi + \frac16 \{\gamma^{\underline{de}},\gamma^{\underline{abc}}\}_{\alpha\gamma}H_{\underline{abc}}\right)\Big|_{\theta=0}=0 
\end{align}
Computing the (anti-)commutators, the above expression together with (\ref{closedSusy}) implies that 
\begin{align}
    H_{a\underline{bc}}|_{\theta=0}=0.
\end{align}
It would also be possible to require an $AdS$ super algebra from (\ref{closedSusy}) by requiring it to be proportional to a four dimensional Lorentz rotation, however it is not possible to have four dimensional $AdS$ solutions without including a gaugino condensate \cite{LopesCardoso:2003sp,Frey:2005zz}. Furthermore, without considering a warp factor in the four dimensional metric, all $H$ flux has to vanish. In this particular case, for spinors satisfying (\ref{singlets}) the equation (\ref{lasteq}) will imply that 
\begin{align}
    R_{a\underline{bcd}}=R_{IJ\underline{cd}}=R_{\bar I\bar J\underline{cd}}=\delta^{I\bar J}R_{I\bar J\underline{cd}}=0.
\end{align}
Finally, these conditions together with \ref{InitialCondSusy} and  Bianchi identities imply the four dimensional space is flat and the internal space is Ricci flat and K\"ahler.

\vskip 0.3in
{\bf Acknowledgements}~
O{\sc c} and B{\sc cv} would like to thank William D. Linch {\sc III} for discussions and commments on the draft and Fondecyt grants 1200342 and 1201550 for partial financial support.
\vskip 0.3in

\appendix

{
\bibliographystyle{abe}
\bibliography{mybib}{}
}
\end{document}